# Tetrathiomolybdate Modified Au Electrodes: Convenient Tuning of the kinetics of Electron Transfer and its application in Electrocatalysis

Sudipta Chatterjee,[†] Kushal Sengupta,[†] Abhishek Dey*

Department of Inorganic Chemistry, Indian Association for the Cultivation of Science, Jadavpur, Kolkata 32.

*To whom all correspondence should be done

**Abstract**

Ammonium tetrathiomolybdate (ATM) spontaneously self assembles on Au electrodes forming a hydrophilic, air stable, pH (3-11) tolerant multilayer that is stable over a reasonably large potential window. The ATM functionalized Au electrodes can adsorb iron porphyrin catalysts and act as $O_2$ reducing electrodes. These electrodes are stable enough to perform rotating disk electrochemistry (RDE) as well as rotating ring disk electrochemistry (RRDE) experiments. The X-ray photoelectron spectroscopy (XPS) data indicate that the sulphide atoms of ATM anchors a single ATM layer on to Au and the subsequent layers grow vertically due to the presence of hydrogen bonding $NH_4^+$ counter-ions. The formation and growth of these ATM adlayers is investigated using atomic force microscopy (AFM), scanning electron microscopy (SEM) and a series of electrochemical data. The ATM functionalized Au electrodes have double layer capacitance comparable to those reported for Au electrode bearing short chain alkyl thiol self assembled monolayer (SAM). Importantly, the rate of interfacial charge transfer (CT) can be tuned by controlling the thickness of the adlayers by simply adjusting the deposition time. Importantly the kinetics of a catalyst adsorbed on this ATM adlayer can be switched from mass transfer limited to CT limited regime by adjusting the deposition time.

## 1. Introduction

Modifications of electrode materials with a wide range of substrates have become a useful tool for sensing and electrocatalysis.[1-5] Self-assembly of a molecule on a substrate requires spontaneous organization of molecules at the solid-liquid interface. This leads to the inhibition of inherent electrochemical response of the bare electrode surface and formation of either a well ordered architecture or an aggregation of entities with newer functions.[6, 7] Several instances of self-assembly of organic moiety on metals (like Au, Ag, Cu, Pt, Hg),[8-13] non-metals (like graphite, silicon, graphene),[14-17] metal oxides (like ITO, PTO, $Al_2O_3$, $Fe_xO_y$, $Ti/TiO_2$),[13, 18-22] nano-structured materials[23-25] and semi-conductors[26, 27] are reported. These organic moiety have a terminal N-, O- and S-atom which link to the electrode material.[28-33] Among them the Au-S interface is a thriving area of research because of their easy assembly and strength of the Au-S bond. In particular, thiol modified Au surfaces have proved to be very useful for the attachment of electroactive species as the terminal head groups of these linkers can be easily modified with different functional groups like -$CO_2H$, -OH, -SH, -$NH_2$, -$N_3$, -CN etc.[6, 34-47] Such modification have allowed controlling the microenvironment of the electrode surface and site specific attachment of synthetic and biological moieties.[24, 42, 48-52] Additionally, the rate of interfacial charge transfer (CT) can be tuned by controlling the chain length of the thiols.[53] This has been widely demonstrated by Chidsey where the CT rates could be lowered from $10^4$ $s^{-1}$ to 6-10 $s^{-1}$ by increasing the chain length of the thiol from octane thiol to hexadecane thiol.[6, 54] In spite of some undesirable drawbacks (e.g. lack of stability in organic solvents, high temperature *etc.*) self assembled monolayer (SAM) of organothiols on Au, Ag, and other metal surfaces remain one of the most versatile electrode modification techniques till date.[55-60]

Physiadsorption or attachment of biomolecules or electrocatalysts on electrodes finds immense application in biomedicine, bioelectronics, biosensors and fuel cells.[61-65] Oxygen reduction reaction (ORR) is a key cathodic reaction in industry where an efficient catalyst is desired that can provide high current density with low overpotentials.[66-69] Several groups have reported the immobilization of porphyrin complexes, either by physiadsorption or by covalent attachment, on electrodes modified with organic thiol SAMs.[42, 51, 70-76] The hydrophobicity of the SAM allows attachment of water insoluble catalysts with organic backbone on to metallic electrodes. These strategies provide a heterogenous platform for studying the ORR properties of water insoluble porphyrins in an aqueous medium.[4, 5, 42, 70, 76, 77] In the recent past, a spectro-electrochemical technique has also been developed where the intermediates involved during the ORR by synthetic iron porphyrins, physiadsorbed on alkanethiol SAM, could be observed directly.[78]

Recently, we have shown that ammonium tetrathiomolybdate $(NH_4)_2[MoS_4]$ (ATM) spontaneously assembles on an Au surfaces. These assemblies generate $H_2$ from water between pH 4-9 with only 20 mV onset potential vs RHE.[79] Nano crystalline sulfides of Mo e.g. $MoS_2$, $MoS_3$ etc. and inorganic Mo-S clusters on gold surface in both neutral and charged forms, owing to its easy availability, and low cost, had already been explored as electrode materials for reduction of $H^+$ to $H_2$.[80-86] Many of these materials are derived from ATM by repeated cathodic and anodic deposition.[80-82, 87, 88]

In this study, we report the details of the spontaneous self-assembly of ATM which could be achieved by immersion of Au surface in an aqueous solution of ATM under ambient conditions. The resultant surfaces are characterized with X-ray photoelectron spectroscopy (XPS), atomic force microscopy (AFM), scanning electron microscopy (SEM) and a combination of electrochemical tools. The surface morphology of ATM self assembly on Au are examined using these spectroscopic and electrochemical methods as a function of concentration of deposition solution and immersion time. These surfaces are quite stable over a wide pH range as well as in several common organic media. These surfaces can be used to immobilize electrocatalysts via physiadsorption and the rate of electron transfer from the electrode can be controlled by tuning the thickness of the ATM adlayers..

## 2. Experiment details

### 2.1. Materials

All reagents were of the highest grade commercially available and were used without further purification. Sodium molybdate pentahydrate $(Na_2[MoO_4].5H_2O)$, and potassium hexaflurophosphate $(KPF_6)$ were purchased from Sigma-Aldrich. Disodium hydrogen phosphate dihydrate $(Na_2HPO_4. 2H_2O)$, potassium chloride (KCl), hydrochloric acid (HCl), aqueous $NH_3$ (98%), and ethanol were purchased from Merck. Sodium sulphide nonahydrate $(Na_2S. 9H_2O)$ was purchased from Rankem, India. Au wafers were purchased from Platypus Technologies (1000 Å of Au on 50 Å of Ti adhesion layer on top of a Si (III) surface).

### 2.2. Instrumentation

All electrochemical experiments were performed using a CH Instruments (model CHI710D Electrochemical Analyzer). Biopotentiostat, and electrodeswere purchased from CH Instruments. The rotating ring disk electrochemical (RRDE) set up was obtained from Pine Research Instrumentation (E6 series ChangeDisk tips with AFE6M rotor). The AFM data were recorded in a Veeco dicp II (Model no: AP-0100) instrument bearing a phosphate doped Si cantilever (1-10 ohm.cm, thickness 3.5-4.5 μm, length 115-135 μm, width 30-40 μm, resonance frequency 245-287 KHz, elasticity 20-80 N/m). The surface morphology of the assembled layers were observed through a field-emission scanning electron microscope (FE-SEM, JSM-6700F), purchased from JEOL LTD, JAPAN. X-ray Photoelectron Spectroscopy (XPS) data were collected using an instrument from Omicron Nanotechnology GmbH, Germany (serial no.-0571).

## 2.3. Synthesis

### 2.3.1. Ammonium tetrathiomolybdate (ATM)

ATM was prepared from $Na_2MoO_4.2H_2O$ according to reported procedure.[89] $Na_2MoO_4.2H_2O$ was dissolved in a 3:1 (by volume) mixture of conc. $NH_4OH$ and $H_2O$. $H_2S$ (generated by dropwise addition of 6 N HCl on pure $Na_2S.9H_2O$) was bubbled through the ammoniacal solution until it was saturated at ambient temperature. The reaction mixture was warmed to about 60 °C for 30 min maintaining a constant flow of $H_2S$. The mixture was then cooled to 4 °C and kept for 30 min. Red crystals precipitated out which was filtered was washed with cold water and ethanol. The product was finally dried under vacuum.

### 2.3.2. α₄-meso-tetra(2-(4-ferrocenyltriazol)phenyl)porphyrinato iron (III) (α₄-FeFc₄)bromide

The α₄-FeFc₄ complex was synthesized as reported in literature.[5]

## 2.4. Construction of Electrodes: Formation of self assembled layer

Au wafers were cleaned electrochemically by sweeping several times between 1.5 V to -0.3 V in 0.5 M $H_2SO_4$. Depositing solutions of 1 mM concentration were prepared by simply dissolving ATM crystals in triple distilled water. Freshly cleaned Au wafers were rinsed with triple distilled water, purged with $N_2$ gas and immersed in the solutions for self assembly. Note that different concentrations of ATM solution have also been used in few experiments and similar procedure has been followed for self assembly. The deposition time has been varied according to requirement and has been clearly stated in the respective experiments. Various surface characterization techniques like XPS, FE-SEM *etc.* have been used where 1 mM ATM solution deposited for 40 min has been used until otherwise mentioned.

**2.5. Characterization of the modified surfaces**

*2.5.1. Atomic Force Microscopy (AFM)*

Freshly cut Au wafers were taken for each AFM analysis where ATM modified surfaces were made as described in section 2.4. The surfaces were thoroughly rinsed with triple distilled water before analysis. AFM data were obtained at room temperature in a Veeco dicp II instrument bearing a phosphate doped Si cantilever (1-10 ohm.cm, thickness 3.5-4.5 μm, length 115-135 μm, width 30-40 μm, resonance frequency 245-287 KHz, elasticity 20-80 N/m).

*2.5.2. Field Emission Scanning Electron Microscopy (FE-SEM)*

Samples were prepared in a similar way like AFM. The surfaces were dried at room temperature and were then observed through FE-SEM applying an accelerating voltage of 5 kV after the surfaces were platinum coated. For all the FE-SEM experiments a working distance of 8 mm was used.

*2.5.3. Contact angle measurement*

Contact angles were measured with distilled water on the modified surfaces using captive drop technique on a goniometer constructed in our institute. Reported contact angles were average values from 3-4 droplets placed on different part of the surfaces. The deviations during the measurements were within $\pm 2^0$.

*2.5.4. X-ray Photoelectron Spectroscopy (XPS)*

XPS samples were prepared by immersing an Au wafer into 1 mM aqueous solution of ATM. The sample was mounted on a standard sample holder for XPS analysis. The sample was then placed into the XPS vacuum introduction chamber and pumped prior to introduction into the main ultra-high vacuum system where the base pressure of the chamber initially was $1 \times 10^{-10}$ mbar and during the experiment was ~3 $\times 10^{-10}$ mbar. XPS uses monochromatic Mg Kα radiation (1253.6 eV) for excitation. High resolution scans, with a total energy resolution of about 1.0 eV, were recorded with pass energy of 20 eV and step size of 0.05 eV. Binding energy spectra were calibrated by the Ag $3d_{5/2}$ peak at 368.2 eV. The area of scan was about 0.5 cm$^2$. An error of ±0.1 eV was estimated for all the measured values. Lorentzian fit has been employed to fit the data obtained for XPS with the help of a standard peakfit software.

**2.6. Physiadsorption of FeFc$_4$ catalyst on the self assembled ATM layer**

FeFc$_4$ was dissolved in chloroform to make a 1 mM solution. ATM modified surfaces were prepared as described in section 2.4. The wafers were taken out of the depositing solutions, washed thoroughly with triple distilled water and dried before every experiment. These were then inserted in the plate testing material. The chloroform solution of the catalyst was added over the surfaces after rinsing with triple distilled water and ethanol and then the catalyst solution was allowed to physiadsorb for 40 min. After 40 min, the surfaces were rinsed with chloroform followed by ethanol and triple distilled water.

**2.7. Electrochemical measurements: Cyclic Voltammetry (CV)**

All CV experiments were done in pH 7 buffer (until otherwise mentioned) containing 100 mM Na$_2$HPO$_4$.2H$_2$O and 100 mM KPF$_6$ (supporting electrolyte) using Pt wire as the counter electrode and Ag/AgCl as the reference electrode. All electrochemical experiments were performed under ambient conditions.

**2.8. Concentration and time dependence studies**

Au wafers were cleaned as mentioned in section 2.4. Separate surfaces were used for each experiment. During concentration dependent studies the wafers were kept in the 0.01 mM, 1mM, and 100 mM depositing solutions of ATM for 30 minutes. During time dependent studies the wafers were immersed into 1 mM solution of ATM for 10 min, 20 min, 30 min, 40 min *etc.*, respectively. The wafers were then taken out and rinsed with triple distilled water and dried under N$_2$ atmosphere. These surfaces were then subjected to electrochemical and AFM studies.

**2.9. Reductive desorption**

Reductive desorption of Au surfaces modified with ATM was done in the ethanolic solution of 0.5 M KOH (saturated with Ar) in a N$_2$ glove box using Ag/AgCl and Pt-wire as the reference and counter electrode respectively and at 20 mV/s scan rate within the potential range of -0.2 V to -1.2 V.

**3. Results and discussion**

**3.1.** *Formation and characterization of the Self Assembled ATM layer*

Clean Au wafers show an O$_2$ reduction current at around -0.35 V (*vs.* Ag/AgCl) in air saturated pH 7 PO$_4^{3-}$ buffer solution during a cyclic voltammetry experiment. However, these clean Au wafers do not show any O$_2$ reduction under similar conditions at these potentials (Figure 1) after keeping immersed in the depositing solution of ATM for moderate time. This clearly indicates the shielding of the Au electrode by ATM. The contact angle value for ATM modified Au surface is found to be $(57\pm1.5)^o$ which is much lower than that of bare Au electrode $(84\pm2)^o$ (Supporting Information, Figure S1). This validates the formation of either a thin film or an assembly which is hydrophilic in nature.[50, 90, 91] Formation of self assembly using a dilute solution generally gives ordered and densely packed monolayers for organic molecules depending on chain length whereas variation in concentration and time of depositing solution leads to the formation of either multilayers or non-uniform scattered layers over the surface.[92-96] Thus, the concentration of the depositing solution and the time of deposition were optimized for the self assembly.

**3.1.1.** *Concentration dependent studies*

To evaluate the dependence of concentration on the assembly of ATM over the Au surfaces three concentrations, 0.01 mM, 1 mM, and 100 mM, of the depositing solution have been used keeping the deposition time same (ca. 30 min). All these cases resulted in the formation of the assembled adlayers as can be observed from the absence of $O_2$ reduction current in their respective CVs at cathodic potentials in an aerated pH 7 buffer solution (Figure 1).

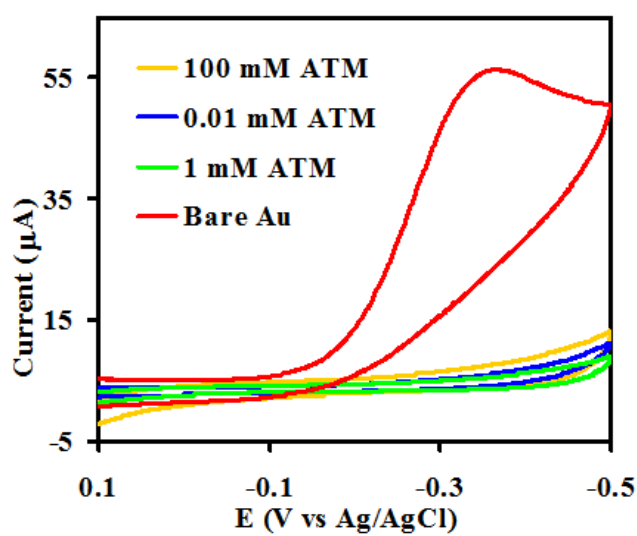

**Figure 1.** CV data of bare Au wafer (red) and the modified Au surfaces with 0.01 mM ATM (blue), 1 mM ATM (green), and 100 mM ATM (yellow) in air saturated pH 7 buffer at a scan rate of 50 mV/s using Ag/AgCl as reference and Pt wire as counter electrodes respectively.

The AFM images of ATM modified Au clearly suggest that the morphology of the layers vary with the concentrations of the deposition solution used. Figure 2 shows the assembly of ATM on smooth Au surfaces (with same cross sectional area in all cases) and the most uniform surface is obtained when 1 mM deposition solution is used (Figure 2B). High resolution FE-SEM data of 1 mM ATM modified surface also shows the homogenous nature of the surface (Figure 2D). The morphology of the bare Au surface is simultaneously measured with the help of AFM (differences in bearing ratio, roughness and height distribution by line analysis) and FE-SEM analyses which show different features in comparison of the ATM modified Au (Supporting Information, Figure S2).[79] The height distribution profiles at different concentrations of ATM show different average vertical lengths ($D_{av}$) *ca.* $D_{av}$=4-6 nm for 0.01 mM, $D_{av}$ = 4 nm for 1 mM, and $D_{av}$= 8 nm for 100 mM of ATM depositing solution (Supporting Information, Figure S3). Note that these values are a direct measure of the thickness of the ATM film on the electrode.[97] A monolayer of ATM would have a height of ~1 nm (as determined from the X-ray crystal structure of ATM)[98] but the average height of the modified surfaces for different concentrations range from 3-5 nm indicating the formation of multiple layers in all these cases. This multilayer formation likely occurs via strong electrostatic and hydrogen bonding interactions between alternated layers mediated by the $NH_4^+$ counter ions present in ATM.

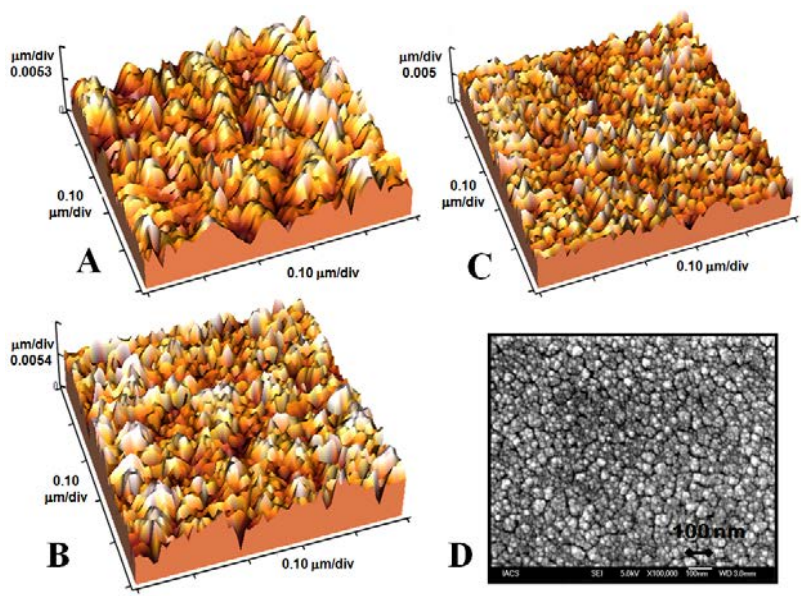

**Figure 2.** AFM images of self assembled ATM layers on Au surfaces prepared at various concentrations of the depositing solution **(A)** 0.01 mM, **(B)** 1 mM, and **(C)** 100 mM. **(D)** FE-SEM image of a similar modified surface prepared from a 1 mM ATM solution.

Electrochemical double layer capacitance ($C_{dl}$) is a useful parameter in understanding the nature of the assembly and the extent of insulation of the electrode surface. In the absence of any redox active species, these assemblies behave like an ideal capacitor. Helmholtz's theory of double layer capacitance predicts that for an ideal capacitor $C_{dl}$ varies with thickness (d) as,

$$C_{dl} = \frac{\varepsilon \varepsilon_0}{d} \quad (1)$$

where, $\varepsilon$ is the dielectric constant of the separation medium and, $\varepsilon_0$ is the permeativity of free space. $C_{dl}$ can be determined from the measured charging current ($I_c$) according to equation 2,

$$C_{dl} = \frac{I_c}{\nu A} \quad (2)$$

where $\nu$ is the scan rate and, $A$ is the geometric area of the electrode surface. The permeability of ionic species through these double layers is dictated by the value of $C_{dl}$. If the electrolyte is impermeable to the assembly, $C_{dl}$ becomes low. However, significant surface defects like pin-hole, non-homogeneity of the surface *etc.* result in high $C_{dl}$ values for modified surfaces.

CV measurements have been carried out on the ATM modified electrodes using different concentration of ATM at a scan rate of 50 mV/s in pH 7 phosphate buffer in the potential window of 0.2 V to -0.2 V vs. Ag/AgCl. Note that, in this region the currents are mainly due to charging of the double layer, and more importantly, this capacitive current is independent of the applied potential, and the electrolyte used. The $C_{dl}$ of the ATM modified electrodes are calculated from the scan rate dependence of charging current density at 0 V vs. Ag/AgCl, where the slope yields the value of $C_{dl}$ (Figure 3A and Supporting Information, Figure S4). The $C_{dl}$ values obtained for the samples prepared from 0.01 mM, 1 mM and, 100 mM depositing solutions are 22±0.5, 13±0.6 and, 32±1 μF/cm$^2$, respectively. Thus the $C_{dl}$ values for 0.1 mM and 100 mM ATM solution modified surfaces are 1.5 and 2.3 times higher than that of the corresponding 1 mM solution (Figure 3C). Hence, a 1 mM depositing solution is for surface modification and has been used for further experiments. The $C_{dl}$ of ATM modified Au is comparable to that obtained for short chain organothiols.[6, 39, 99-101] It is important to note that other amorphous and porous MoS$_x$ (x = 2, 3) films, grown or deposited on the electrode surfaces have $C_{dl}$ values of 2300-2500 μF/cm$^2$ i.e., ~100 times greater than ATM.[81, 102]

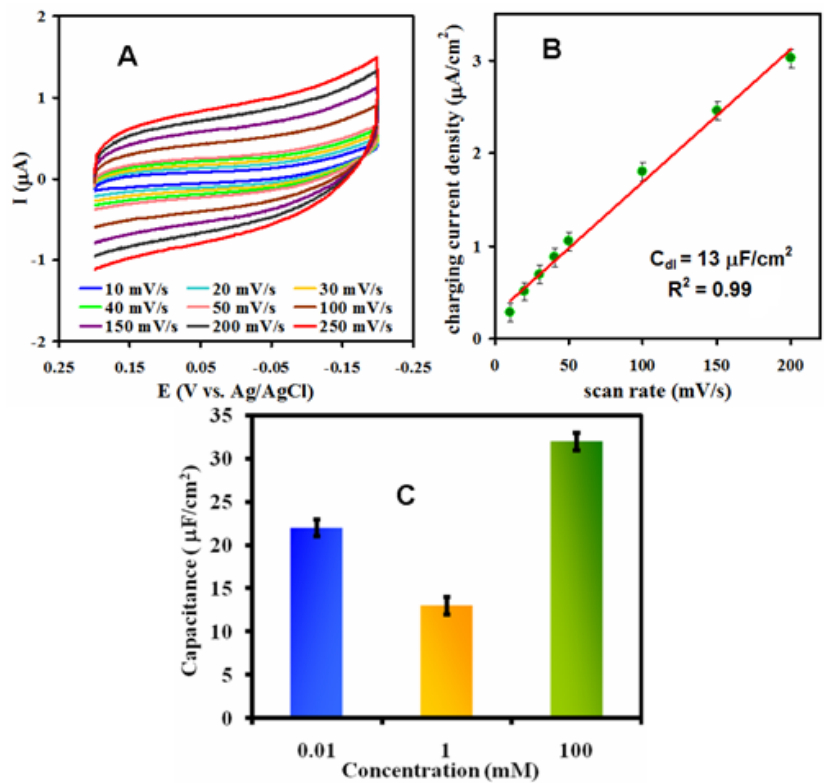

**Figure 3.** **(A)** cyclic voltammograms in the region of 0.2 V to -0.2 V vs. Ag/AgCl and **(B)** scan rate dependence of the charging current density at 0 V potential vs. Ag/AgCl of 1 mM depositing solution of ATM on Au in pH 7; **(C)** A plot of the capacitance values calculated from the CV data obtained at various concentration, 0.01 mM (blue), 1 mM (yellow), and 100 mM (green), of the depositing solution. The scan rate used was 50 mV/s.

### 3.1.2. *Time dependence studies*

The CV data obtained on the modified surfaces at a scan rate of 50 mV/s in pH 7 show a gradual decrease in the charging current density with increase in the time of deposition (Figure 4A). Note that the charging current (at 0V) decreases sharply till 40 min and finally attains a steady value after 180 min (Figure 4B). Time dependent AFM imaging has also been performed to investigate the evolution of the morphology of the ATM modified Au surfaces. The corresponding 2D and 3D topographic views of the modified surfaces, formed by incubating in the 1 mM depositing solution for different time intervals, are shown in Figure 5. From the 2D images it is clearly evident that the bright spots, which are of few nm in diameter, are uniform and closely spaced after 40 min of incubation compared to 10 min of incubation (Figure 5D and E). Although, the average thicknesses of these two self assembled layers do not vary much as reflected in their height distribution profile diagrams (Figure 5G and H), the height is more uniform after 40 min of incubation.

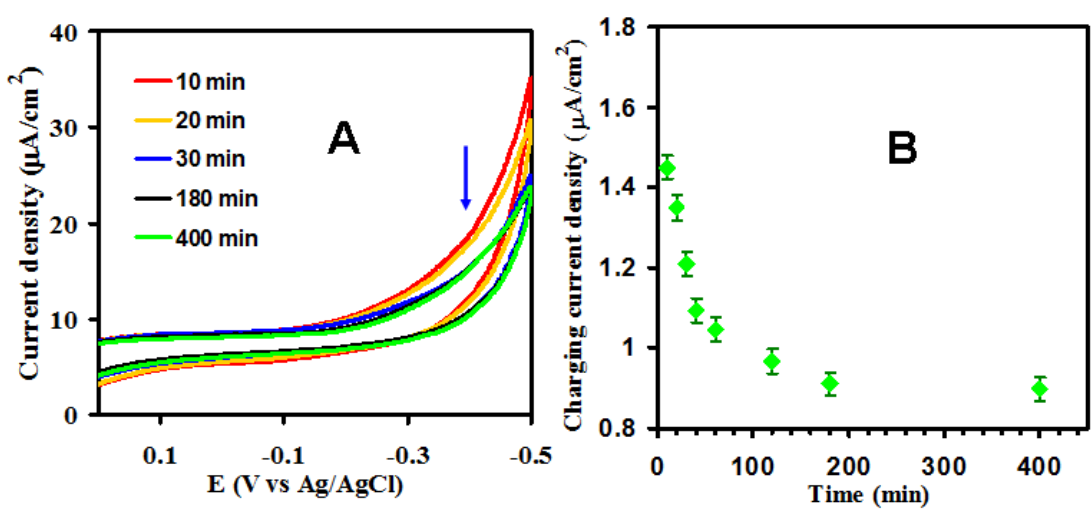

**Figure 4.** **(A)** CV data of 1 mM ATM modified surfaces subjected to various time of deposition in air saturated pH 7 buffer at a scan rate of 50 mV/s using Ag/AgCl as reference and Pt wire as counter electrodes respectively. All the data are not shown. **(B)** A plot of the corresponding charging current values of the respective CV curves as a function of immersion time.

Upon incubating for a longer period of time (400 min), the self assembled layers are found to be non uniform in nature both in the 2D and 3D images (Figure 5). Mainly two types of structures are observed which vary both in width and height (Figure 5C and F). The average heights of the comparatively smaller structures range between 6-8 nm whereas those of the larger structures are between 13-15 nm (Figure 5C, green circle and Figure 5I). The 3D images of these surfaces on an equal cross-sectional area imply the same. Therefore, the AFM images not only suggest the multilayer deposition of ATM resulting from longer incubation of Au surfaces in ATM solution, but also it confirms that assembly obtained around 40 min is more uniform, well-packed and homogeneous compared to ATM assemblies deposited for 10 min and 400 min. Therefore, from the above analyses, it can easily be conceived that 1 mM depositing solution and 40 min of deposition time are the desired ones among the three concentrations and time of incubations used for the depositing ATM solution.[103] Thus during the characterization, coverage calculation and stability experiments of these surfaces in general a concentration of 1 mM ATM solution and 40 min incubation was used for the surface modification.

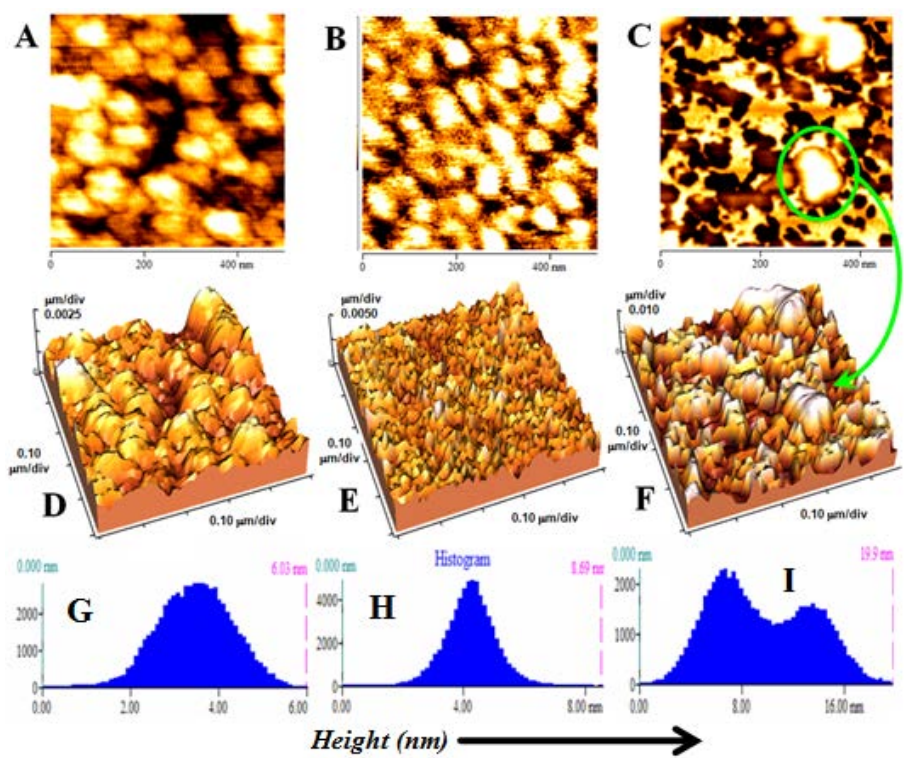

**Figure 5.** AFM images of ATM assembled surfaces. **(A), (B),** and **(C)** are the 2D topology images; **(D), (E),** and **(F)** are the 3D topology images of modified surfaces and **(G), (H)** and **(I)** are the height distribution profiles of the self assembled layers of ATM incubated for 10 min, 40 min and 400 min, respectively.

### 3.1.3. *Heterogenous electron transfer*

The charge transfer through the ATM adlayers can also be characterized electrochemically by the cyclic voltammetric response obtained using water soluble, reversible, one-electron redox agent $Fe(CN)_6^{-3/-4}$. The extent of splitting of the cathodic and anodic peaks ($\Delta E_p$) during $Fe(CN)_6^{3-/4-}$ CV experiment probes the shielding of the electrode. This splitting is sensitive to the thickness of the adlayer and often probes structural defects on the modified surfaces which often lead to improper shielding of the electrodes. The blue line in Figure 6 shows current-potential (*I-E*) response for bare gold with 1 mM $Fe(CN)_6^{-3}$ as the electroactive species in pH 7 phosphate buffer using $KPF_6$ as the electrolyte at a scan rate of 50 mV/s. The splitting between the cathodic and anodic peaks is found to be 150 mV. This indicates that the ferricyanide reduction process is diffusion limited. For an ATM covered Au surface, the $\Delta E_p$ gradually increases with deposition time. The $\Delta E_p$ recorded after 10 min is 170 mV which continued to increase to 410 mV after 180 min of deposition time (Figure 6A). No further increase in $\Delta E_p$ was observed with time (Figure 6B and Supporting Information, Table S1). These results reflect the growth of the ATM adlayers with time, indicating that after about 180 mins of depositing time there is no significant increase in the thickness of the ATM layers.

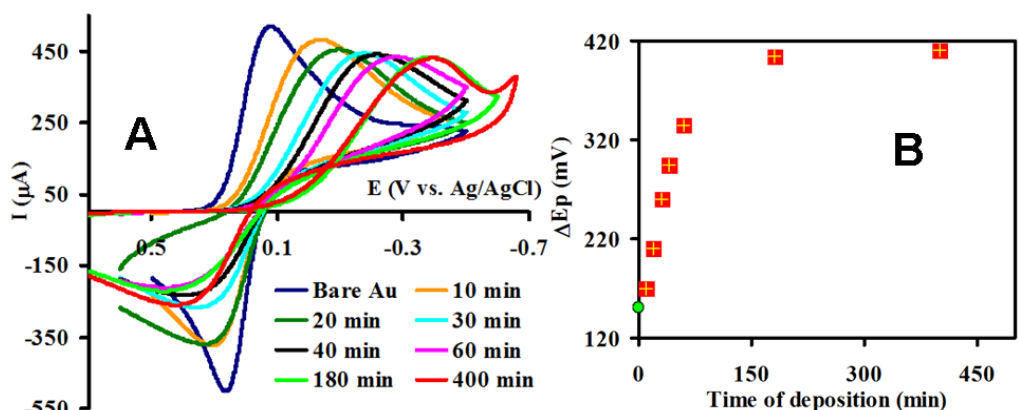

**Figure 6.** **(A)** CV data of 1 mM ATM modified surfaces with different incubation time in pH 7 buffer containing 10 mM $K_3[Fe(CN)_6]$ using Ag/AgCl as reference and Pt wire as counter electrodes respectively; **(B)** A plot of $\Delta E_p$ vs time of deposition of the corresponding CVs obtained in figure 6A. The green point in figure 6B corresponds to the bare electrode i.e. in the absence of any ATM on the Au surface.

Thus the charge transfer to the electroactive species in solution is inhibited as the ATM adlayer on Au grows with time. In the case of organic SAM the increase in chain length of the thiols makes the electron transfer sluggish until the cathodic and anodic peaks slowly dwindle. Unlike an organic SAM, in the ATM self assembly, complete disappearance of the $Fe(CN)_6^{-3/-4}$ CV does not occur. This is likely due to the fact that the ATM assembly is an ionic assembly on the electrode. Hence this assembly will always allow CT; albeit at variable rates.

### 3.1.4. *X-Ray Photoelectron Spectroscopy (XPS)*

XPS provides a useful tool to elucidate the chemical composition of the self assembly of ATM. It helps to determine the oxidation states and nature of bonding of the elements that are closed to these modified surfaces.[83, 104] XPS data of Au electrode bearing layers of 1 mM ATM deposited for 40 min show the Mo $3p_{3/2}$ and $3p_{1/2}$ peaks at 395.4 eV and 413.2 eV; Mo $3d_{3/2}$ and Mo $3d_{5/2}$ peaks at 232.5 eV and 229.7 eV; N 1s peak at 400.0 eV; a major S species (I S) of S 2s, S $2p_{1/2}$ and S $2p_{3/2}$ peaks at 225.8 eV, 163.0 eV, and 161.6 eV; a minor S species (II S) of S 2s, S $2p_{1/2}$ and S $2p_{3/2}$ peaks at 226.9 eV, 164.5 eV, and 162.3 eV; O 1s peak at 532.5 eV (Supporting

Information, Figure S5), respectively.[105] The major $S_{2s}$, $S2p_{1/2}$ and $S2p_{3/2}$ ionization peaks originate from uncoordinated sulfide ion of the ATM assembly.[106] The 2nd set of $S2s$, $S2p_{1/2}$ and $S2p_{3/2}$ ionizations are higher in energy and originate from a sulfide ligand with higher effective nuclear charge (i.e. deeper core orbital energies) than that of the free sulfide of $MoS_4^{2-}$. This is likely due to the coordination of Au to the sulfide ions at the surface. The characteristic peaks for Mo, N, and O confirm the presence $Mo^{VI}$, $NH_4^+$ and water in the assembly. A model of ATM assembly on Au surface where the formation of multilayer of ATM attached to the Au surface via a Mo-S-Au linkage has been recently proposed. These multilayers are stabilized by three dimensional hydrogen bonding network between the constituent ions of ATM. Moreover, the presence of water molecule in this assembly may facilate the formation of such multilayers.

### 3.1.5. *Reductive desorption*

Reductive desorption experiments provide a means by which the surface coverages of organothiols adsorbed on Au and Ag may be determined.[39, 100, 107] Based on the charge consumed during reductive desorption, the number of adsorbed species per square cm is calculated following the reported procedure. The reductive desorption of 1 mM ATM adlayers deposited for 40 min clearly shows a cathodic peak at -0.85 V (*vs.* Ag/AgCl) from which the average surface coverage can be estimated to be $1.34 \pm 0.04 \times 10^{15}$ molecules/cm$^2$ (Supporting Information, Figure S6). This value of surface coverage is nearly 3-4 times more than the coverages of Au fully covered with alkanethiol monolayers.[51, 108]

### 3.2. *Stability of the self assembly*

The tolerance of the ATM adlayer on Au to a wide pH range was investigated. Organic SAMs having different functionalities have been found to be stable at a pH range 1-12 where surface modification reactions and immobilization of biological or synthetic materials can be done.[109-111] The ATM self assembly is found to be stable within a wide pH range of 3-11 (Figure 7A). The $C_{dl}$ values, measured at 0 V vs Ag/AgCl, in the pH range 3-11 range between 12-13 μF/cm$^2$ (Figure 7A), but, at pH below 3 the capacitance of these surfaces increases to 16 μF/cm$^2$ (as indicated by the red point in Figure 7A). Below pH 2 the capacitance could not be calculated accurately as the surface started degrading showing oxygen reduction current by the exposed Au electrode. This increase in charging current *i.e.* capacitance at lower pHs is probably due to the protonation of the sulphur ligands attached to the Au surface leading to the dissolution of the assembly.

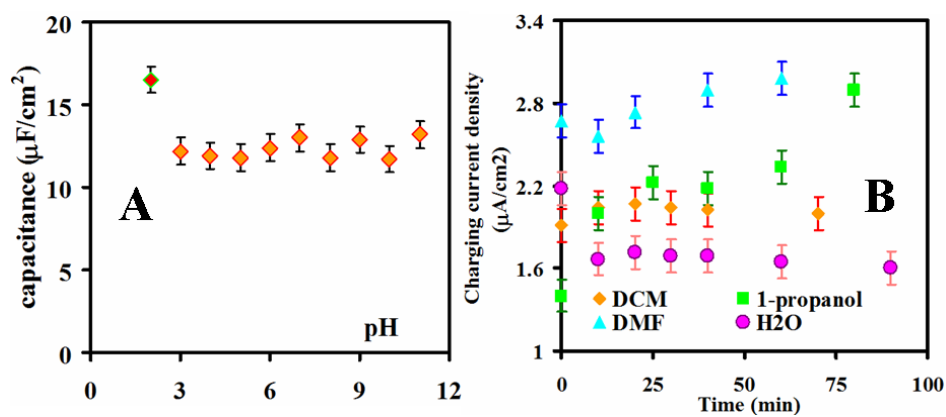

**Figure 7.** (**A**) Plot of double layer capacitance values calculated from the slope values of their corresponding CV data of ATM modified surfaces at different scan rates in different pHs *vs.* pH. (**B**) Plot of charging current density of ATM assembly on Au surfaces as a function of time of exposure to different organic solvents – DCM (orange), 1-propanol (green), DMF (sky blue) and to water as well. The scan rate for this experiment was 50 mV/s.

Self assembled monolayers or bilayers of alkylthiols are known to tolerate a few organic solvents. Likewise, this ATM assembly is found to be stable in few organic solvents like dichloromethane, 1-propanol, dimethylformamide *etc*. (Figure 7B). While this assembly is very stable in DCM, it started degrading with time in 1-propanol and DMF, as can be seen from the corresponding charging current density *vs.* time plot in Figure 7B. Several solvent properties such as polarity, viscosity, solubility as well as interaction of the solvents with both adsorbate and gold surface greatly influence the stability of such self assembled layers.[47, 109, 112]

### 3.3. *Physiadsorption of electrocatalysts for ORR*

We have, in the recent past, reported an iron porphyrn complex bearing four ferrocene (Fc) groups which is an electrocatalyst for ORR. This complex shows selective $4e^-/4H^+$ ORR at pH 7 under both fast and slow electron transfer (ET) conditions. This $FeFc_4$ catalyst can be physiadsorbed on this ATM modified Au surface as indicated by cyclic voltammetric response. A CV around +0.35 V was observed for $Fc/Fc^+$ couple, and a catalytic current at -0.35 V is observed due to oxygen reduction in an aerated pH7 buffer solution (Figure 8). The decrease in $C_{dl}$ and the increase in $\Delta E_p$ of ferrocyanide indicated that the extent of shielding of the Au electrode by ATM could be tuned by adjusting the time. When FeFc$_4$ was immobilized on an ATM assembly which showed the greatest shielding (400 min deposition) of 1 mM solution, the $\Delta E_p$ of the Fc/Fc$^+$ process increased (Figure 8, inset) and the mass transfer limited ORR current is replaced by an ET limited catalytic current (Figure 8). Note that the ORR shifts to more negative potentials as the time of deposition of ATM on Au surfaces increase with concomitant decrease in the corresponding catalytic oxygen reduction current. Thus the ATM assembly on Au allows tuning of ET rates from the electrode to a physiadsorbed species by controlling the deposition time (Figure 8). Chronoamperometric (CA) response of the Fc/Fc$^+$ process of the FeFc$_4$ complex adsorbed on ATM adlayer obtained at different deposition times clearly show a decrease in interfacial ET rates with increase in deposition times (Supporting Information, Figure S7).

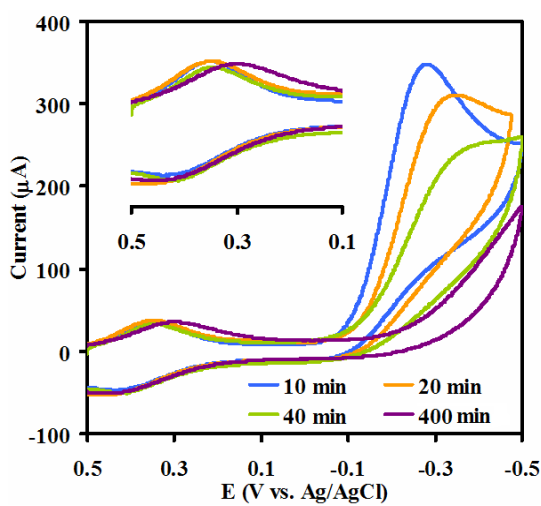

**Figure 8.** Cyclic voltammogram of FeFc$_4$ immobilized on different ATM modified surfaces with different time of deposition - 10 min (blue), 20 min (orange), 40 min (green), and 400 min (violet), in pH 7 buffer at a scan rate of 50 mV/s. Inset shows the zoomed in CV of Fc/Fc$^+$ of the same.

### 3.4.1. *Rotating disc electrochemistry (RDE)*

Electrocatalytic O$_2$ reduction reaction (ORR) at different rotation rates for physiadsorbed FeFc$_4$ on ATM modified Au discs have been performed for different time of deposition of ATM assembly (Figure 9). RDE not only provides the number of electron transferred during ORR but also helps in determining the stability as well as durability of the electrocatalyst as well as that of the ATM adlayer on the electrodes. The O$_2$ reduction current increases with increasing rotation rates following the Koutecky–Levich (K-L) equation, $I^{-1} = i_K^{-1} + i_L^{-1}$, where $i_K$ is the potential dependent kinetic current and $i_L$ is the Levich current (Figure 9A).[113] The K-L analyses yields (details are given in the Supporting Information) that FeFc$_4$ reduces O$_2$ to H$_2$O efficiently by a 4e$^-$/4H$^+$ process in case of ATM assembly deposited for 40 min (Figure 9B). Note that RDE could not be performed on FeFc$_4$ adsorbed on ATM surfaces that have been obtained after 400 min of deposition as these electrodes do not show substrate diffusion limited current any more.

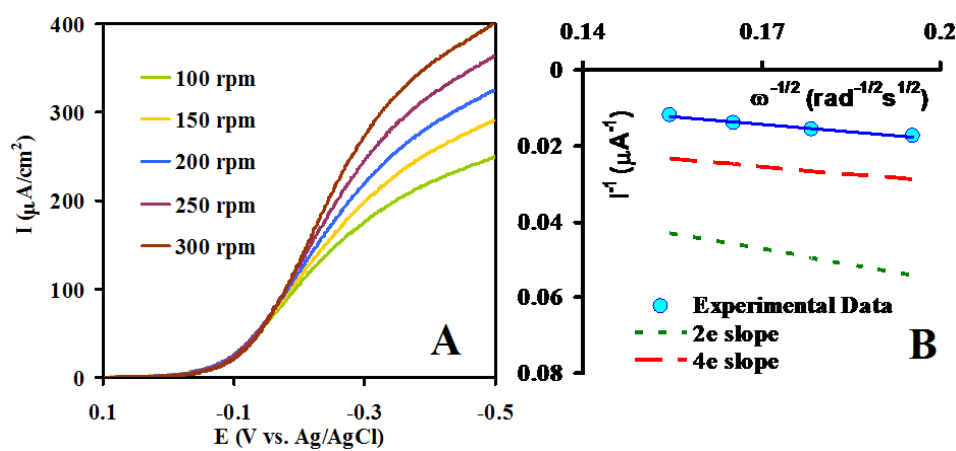

**Figure 9.** **(A)** LSV scans of FeFc$_4$ catalyst physiadsorbed on ATM modified (deposited for 40 min) Au in air saturated pH 7 buffer at a scan rate of 50 mV/s at multiple rotations using Ag/AgCl as reference and Pt wire as counter electrodes. **(B)** K-L plots of the respective catalysts are given in blue bold line in the inset of the figures. The theoretical plots for 2e$^-$ and 4e$^-$ processes are indicated by the dotted (green) and dashed (red) lines respectively.

### 3.4.2. *Rotating ring disc electrochemistry (RRDE)*

Rotating ring disc electrochemistry (RRDE) has been used to calculate the amount of partially reduced oxygen species (PROS) produced due to incomplete reduction of O$_2$. In this technique any O$_2^-$ or O$_2^{2-}$ produced in the modified Au working electrode due to 1e$^-$ or 2e$^-$ reduction of O$_2$ is radially diffused, due to the hydrodynamic current created by the rotation, to the ring, which is held at 0.7 V, where these are oxidized back to O$_2$.[113, 114] This results in an oxidation current in the ring and the ratio of the catalytic current of the ring ($i_r$) and the disc ($i_d$) yields the % of PROS produced. In a generally accepted mechanism generation of PROS entails hydrolysis of a Fe$^{III}$-O$_2^-$ species produced during O$_2$ reduction in the aqueous environments. Note that no current should be detected in the ring if O$_2$ is completely reduced to H$_2$O by the catalyst. When FeFc$_4$ is immobilized on 10 min ATM assembly it produces 10±0.5 % PROS (Supporting Information, Figure S8). However, for 40 min and 400 min assembly it produces 12±1 % and 13±0.5 % PROS (Figure 10). This reproduces previous results obtained with FeFc$_4$ on edge plane graphite (EPG) (ET rate >10$^5$ s$^{-1}$), octanethiol (C$_8$SH) SAM (ET rate ~10$^4$ s$^{-1}$), and hexadecanethiol (C$_{16}$SH) SAM (ET rate ~6-10 s$^{-1}$) electrodes where selective 4e$^-$/4H$^+$ O$_2$ reduction was observed irrespective of ET to the catalyst.[5]

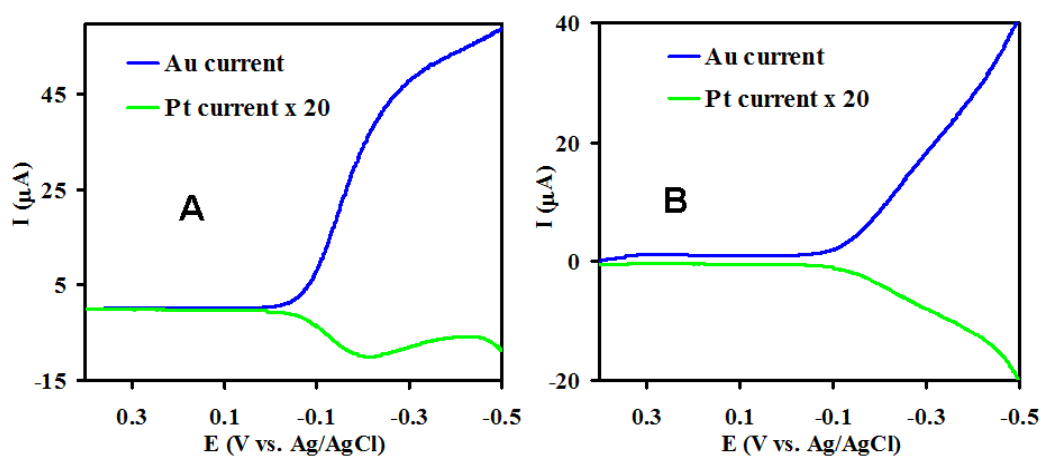

**Figure 10.** RRDE data of FeFc$_4$ physiadsorbed on ATM modified Au electrode, deposited for 40 min (**A**) and 400 min (**B**) showing the disk and Pt ring currents, in air saturated pH 7 buffer at a scan rate of 10 mV/s and rotation speed of 300 rpm, using Ag/AgCl reference and Pt wire counter electrodes.

A combination of spectroscopic and electrochemical techniques indicate the formation of a new hydrophilic surface by simple immersion of clean and smooth Au surfaces into an aqueous solution of ATM. Formation of molecular assembly by enormous number of organic entities has been reported previously where monolayer formation is achieved. Reports suggest an island expansion process which terminates immediately after the formation of single monolayer. Unlike other Mo-S clusters[86, 115] and synthetic inorganic complexes[104, 116] reported earlier where monolayer formation is achieved, this ATM forms multilayer in every cases i.e. when both the concentration as well as time of deposition of the depositing solution has been varied.

The deposition phenomena and growth pattern have been investigated as a function of the concentration of the deposition solution and the time of deposition using AFM, SEM, and CV experiments. These studies establish that 1 mM depositing solution and moderate immersion time (~40 min) yields relatively much uniform multilayer of ATM and minimizes overgrowth among the various conditions (three different concentrations and time of incubations) studied. However, the exact reason or mechanism for the controlled growth with time is not clearly understood yet.

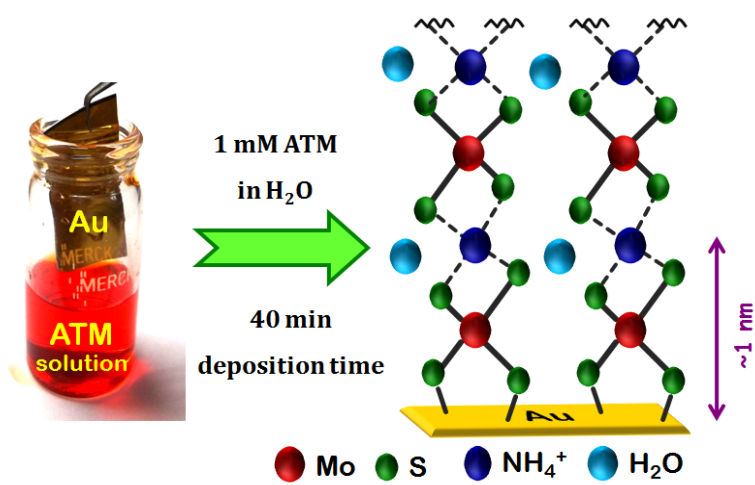

**Figure 11.** Schematic representation of the formation of ATM adlayers on Au surface after 40 min of deposition time. The concentration of ATM used was 1 mM in $H_2O$.

From all the experimental results and analyses above, Figure 11 represents the most likely vertical arrangement of ATM self assembly on Au surface. Since ATM forms multilayer, electrostatic interaction between the ions play a vital role in the stacking phenomena. Note that, the model is only a representation of the assembly and does not necessarily depict the exact conformation or orientation of the species present on the Au surface. The height of the $MoS_4^{2-}$ unit, hydrogen bonded to the counter ammonium ion, is about 1 nm in the crystal structure. The height of the multilayer (4 nm from AFM, Fig. 5H) is consistent with the formation of 4 such layers for 40 min assembly of 1 mM depositing solution of ATM. This can also be explained from the XPS data (*vide infra*).

The model explains the origin of two different sulphur 2s and 2p ionizations in the XPS. One is involved in direct bonding interaction with Au which is reflected in the higher S 2p binding energy. The other S 2p peak, observed at lower binding energy, is due to the unbound sulfidic sulfur of $MoS_4^{-2}$ unit. These sulfur atoms are not directly attached to the Au surface and thus bear negative charges. An approximate relative atom composition of the Au bound $S^{2-}$ and only Mo bound $S^{2-}$ is obtained from the intensities of the $S_{2p}$ peaks to be 1:7, as discussed earlier in our previous report.[79] The above ratio apparently is indicative of the fact that while two sulfides of a $MoS_4^{2-}$ anion binds to the Au, there are 14 unbound sulfides present in the 4 nm high assembly i.e. 2 free sulfide from the bound $MoS_4^{2-}$ and 12 additional sulfides from three assembled $MoS_4^{2-}$ units, in total an assembly of 4 ATM units. The presence of $NH_4^+$ ion in this multilayer assembly is desired as it helps in forming the stacked layer executing electrostatic and hydrogen bonding interactions with sulfur atoms and the intervening water molecules, as shown in Figure 11. These water molecules may serve to mediate the hydrogen bonding network throughout the assembled adlayer.

The ATM adlayers are stable over a reasonable potential, pH and organic solvents. It can be used as a means to immobilize electrocatalysts on Au electrodes as these electrocatalysts physiadsorb on these adlayers. These aslayers are stable enough to allow dynamic electrochemical measurements like RDE and RRDE. A unique attribute of these adlayers is the fact that interfacial CT across these adlayers could be tuned by simply adjusting the deposition time. With increasing deposition time (from 10 min to 400 min) the $C_{dl}$ decreases, $\Delta E_p$ of ferrocyanide increases and $O_2$ reduction by a $FeFc_4$ electrocatalyst becomes ET controlled from mass transfer controlled with increase in deposition time. This control over ET rate by simply varying deposition times cannot be achieved by ordinary alkanethiol SAM.

**Supporting Information**

Contact angle, AFM, FE-SEM, XPS, CV and chronoamperometry data are available in Electronic Supplementary Information.

**Author contribution**

† These two authors contributed equally to this work.

**Notes**

The authors declare no competing financial interest.

**Acknowledgements**

This research is funded by Department of Science and Technology, India, (SR/IC/35-2009) and Department of Atomic Energy, India, (2011/36/12-BRNS). SC and KS acknowledge CSIR-SRF fellowship.

**Table of content graphic:**

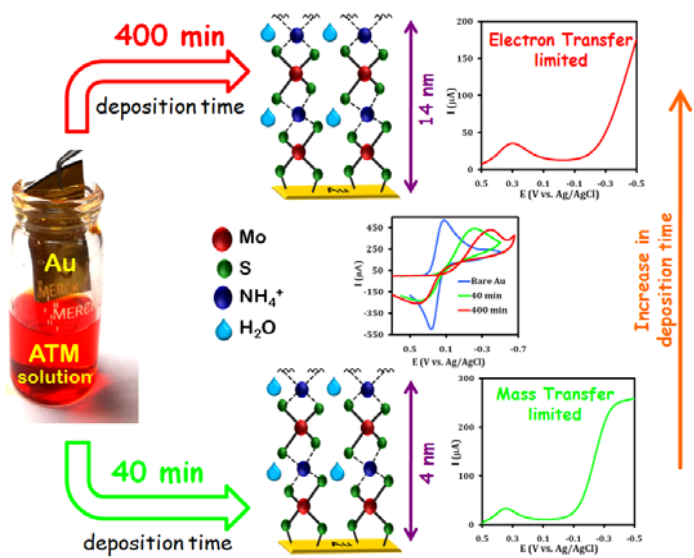

# *Supporting Information*

# *for*

**Tetrathiomolybdate Modified Au Electrodes: Convenient Tuning of the kinetics of Electron Transfer and its application in Electrocatalysis**

Sudipta Chatterjee,[†] Kushal Sengupta,[†] Abhishek Dey*

Department of Inorganic Chemistry, Indian Association for the Cultivation of Science, Jadavpur, Kolkata 32.

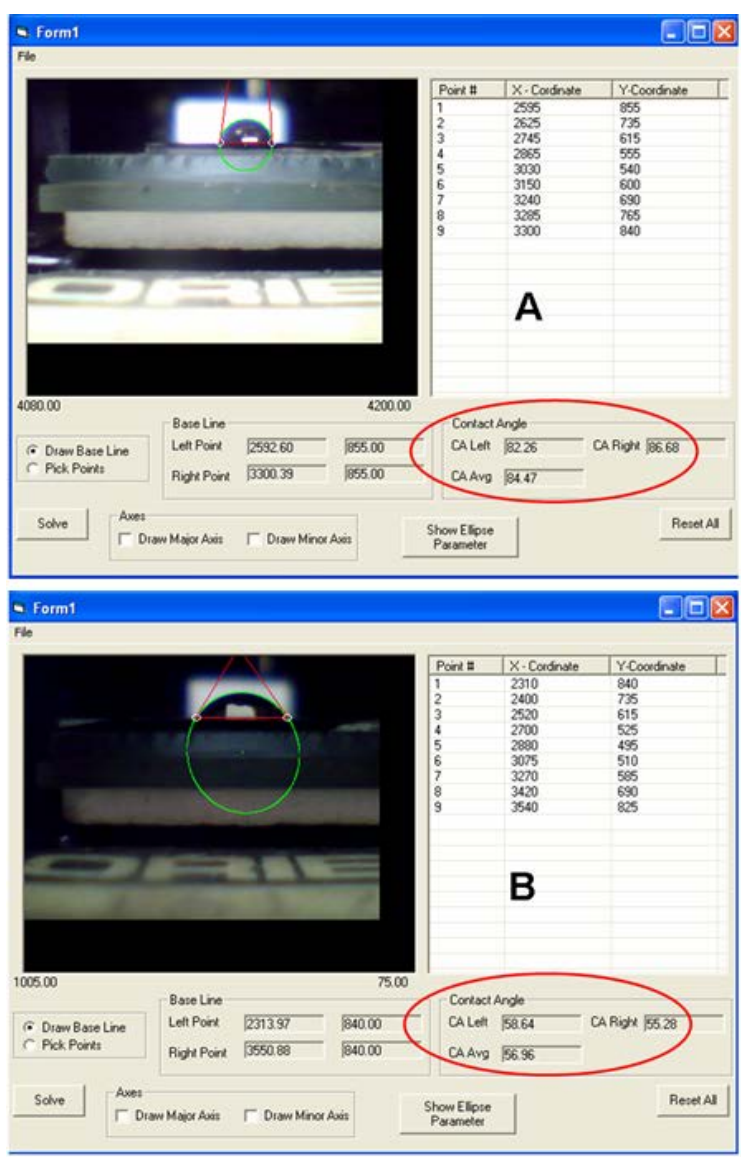

**Figure S1**: Water drop contact angle of (A) bare Au, and (B) ATM deposited for 30 minutes on Au surface.

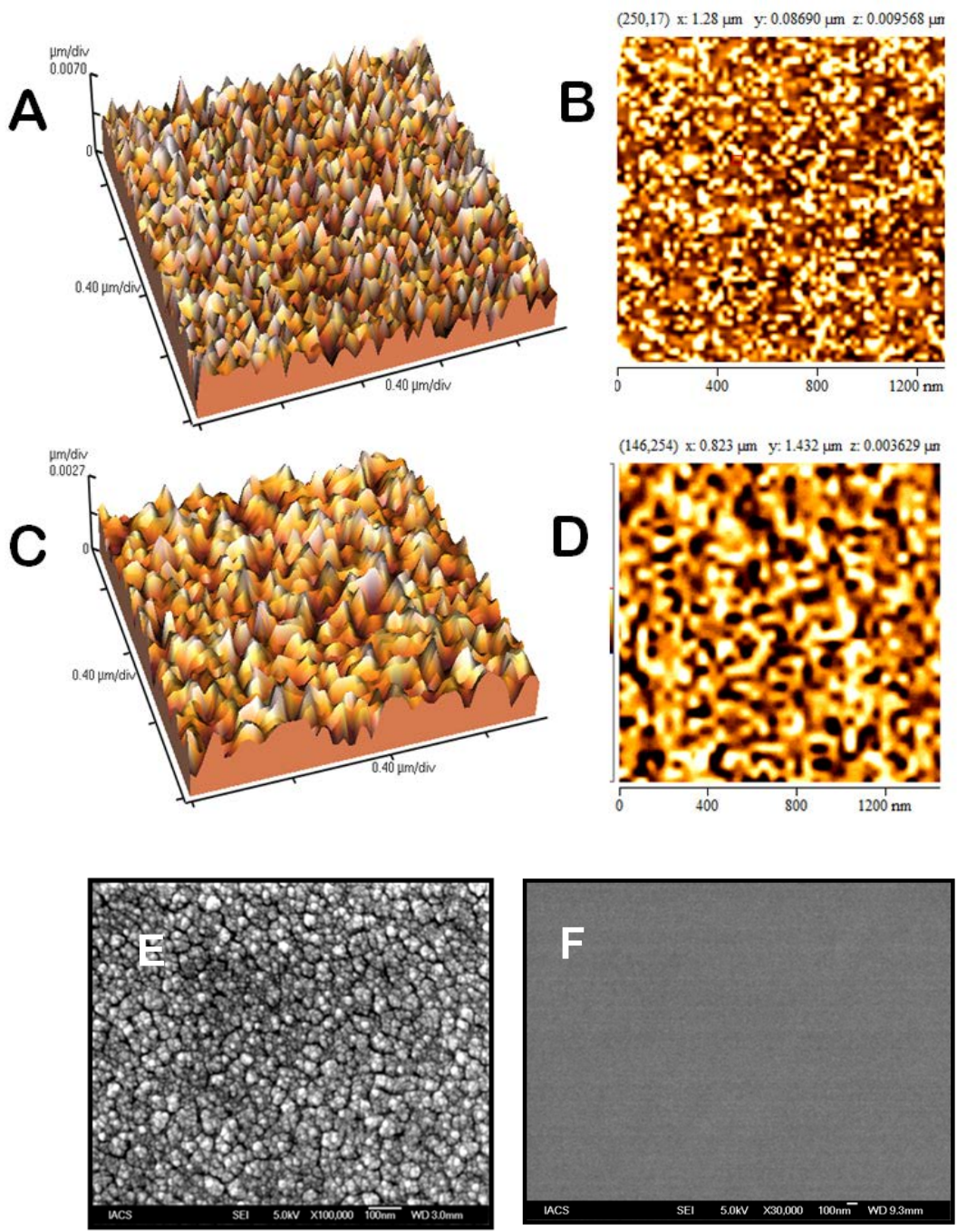

**Figure S2**: 3D and 2D topographic view of ATM modified Au (A and B) and bare Au (C and D), both having same cross sectional area, respectively. FE-SEM image of ATM modified Au (E) bare Au surface (F), respectively.

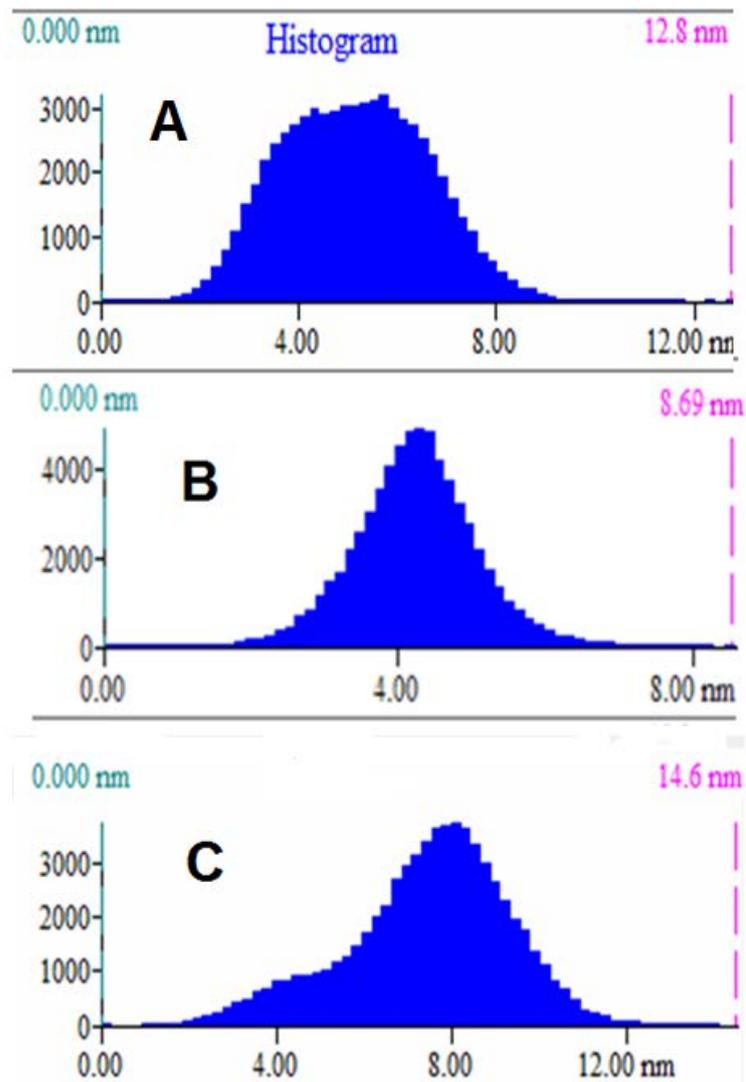

**Figure S3**: Height distribution profiles obtained from AFM topographs of the self assembled layers of (A) 0.01 mM, (B) 1 mM, and (C) 100 mM ATM incubated for 30 mins.

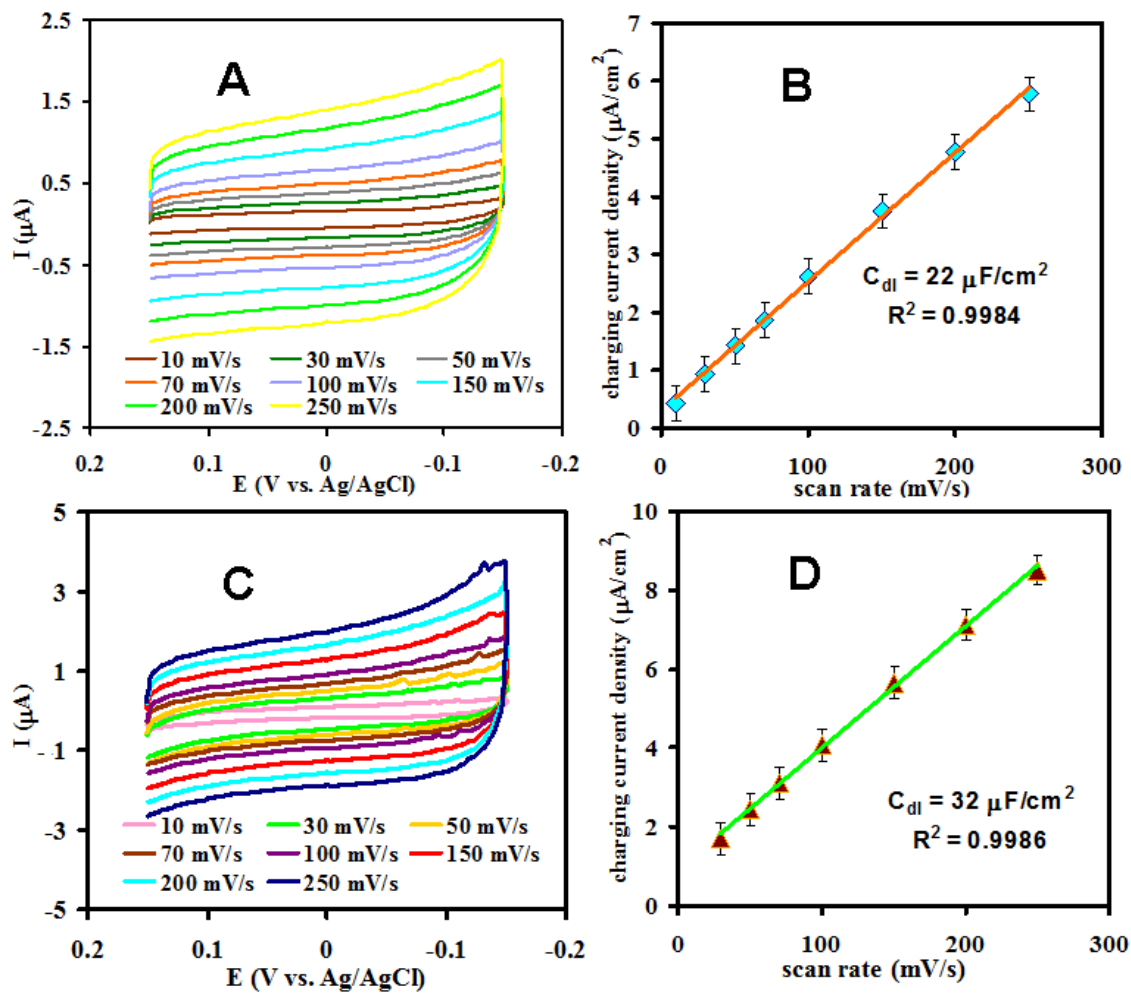

**Figure S4.** (A) and (C) Cyclic voltammograms in the region of 0.15 V to -0.15 V *vs.* Ag/AgCl; (B) and (D) scan rate dependence of the charging current density measured at 0 V potential *vs.* Ag/AgCl of 0.01 mM and 100 mM depositing solution of ATM on Au in pH 7, respectively.

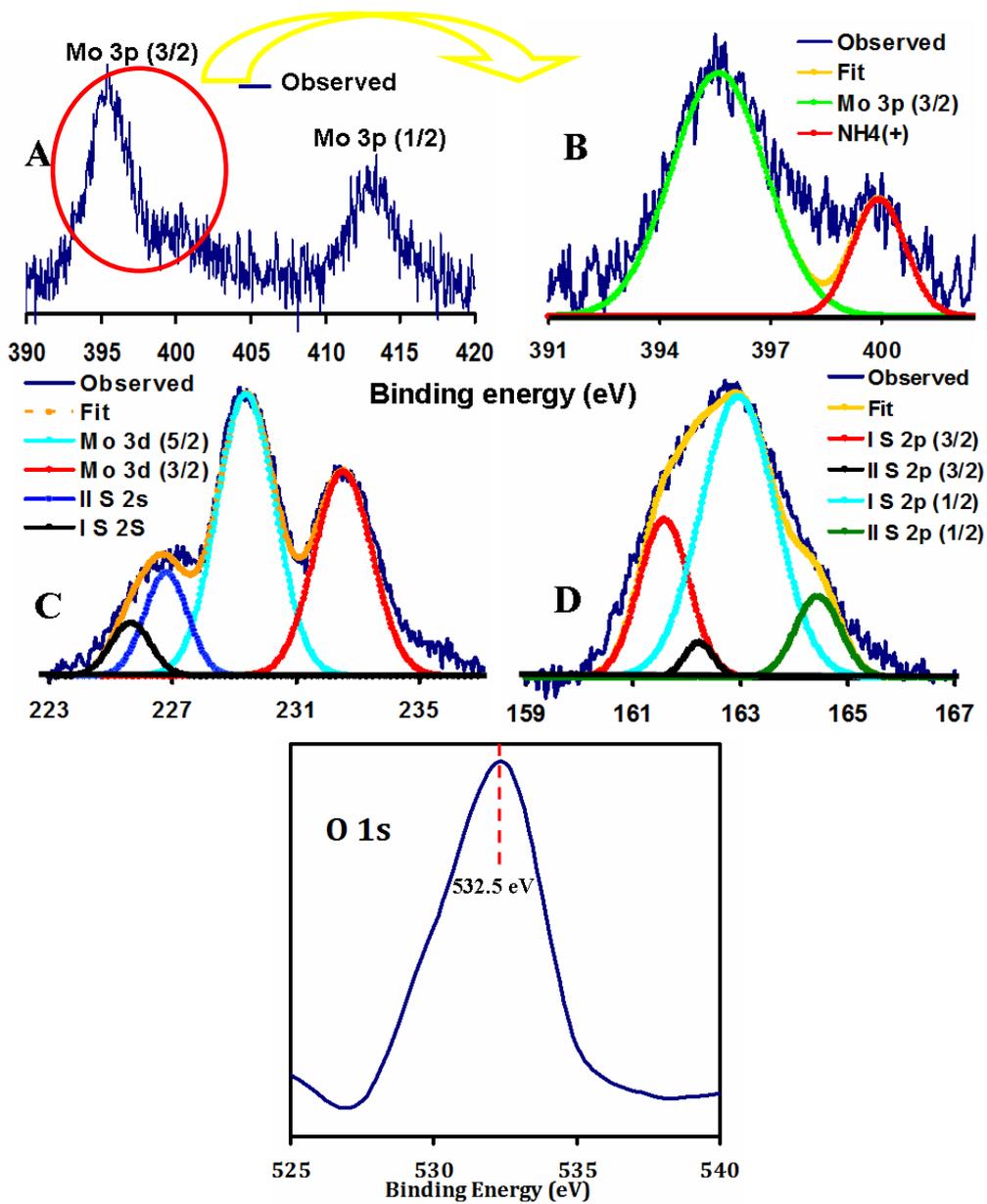

**Figure S5:** XPS data of an ATM modified surface. (A) The Mo 3p region, (B) enlarged view of (A) showing Mo 3p$_{3/2}$ along with NH$_4^+$, (C) the Mo 3d and S 2s region, (D) the S 2p region along with the different components, and (E) O 1s region.

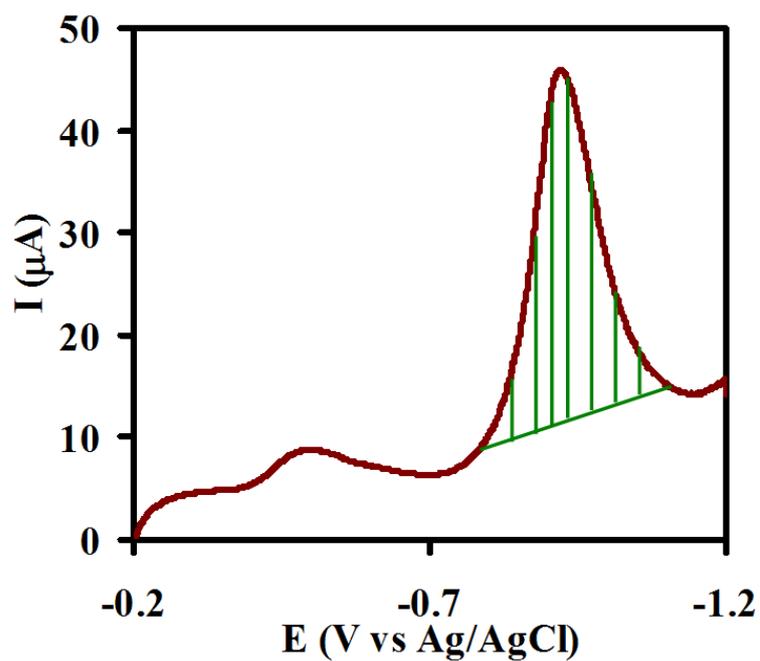

**Figure S6.** The curve represents linear sweep voltammogram of 1 mM ATM adlayer deposited for 40 min on Au in 0.5 M KOH. The scan rate was 20 mV/s. The potential is swept between -0.2 and -1.2 V. Pt and Ag/AgCl were used as counter and reference electrodes, respectively. The marked area indicates the area integrated in order to calculate the surface coverage.

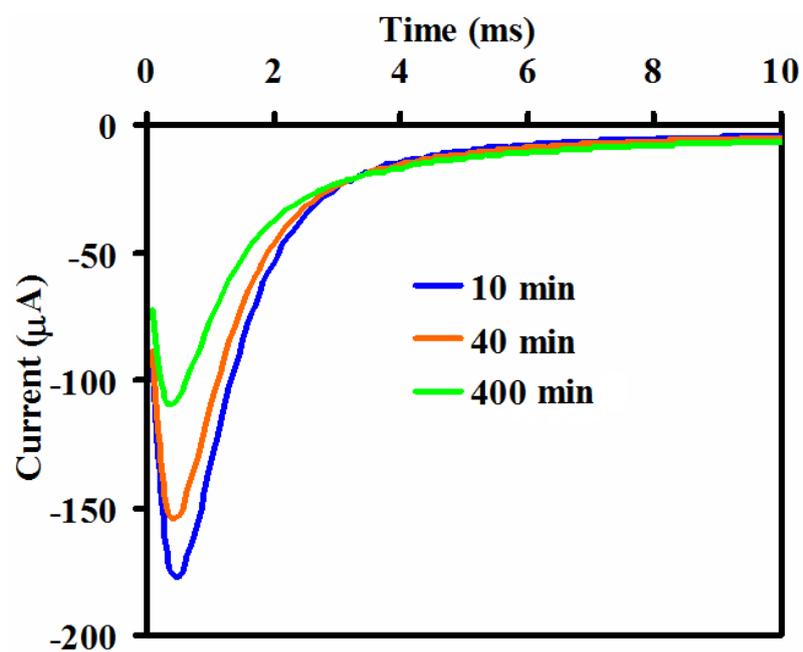

**Figure S7**. Chronoamperometric response of FeFc$_4$ catalyst deposited on Au surfaces modified with ATM for different time interval.

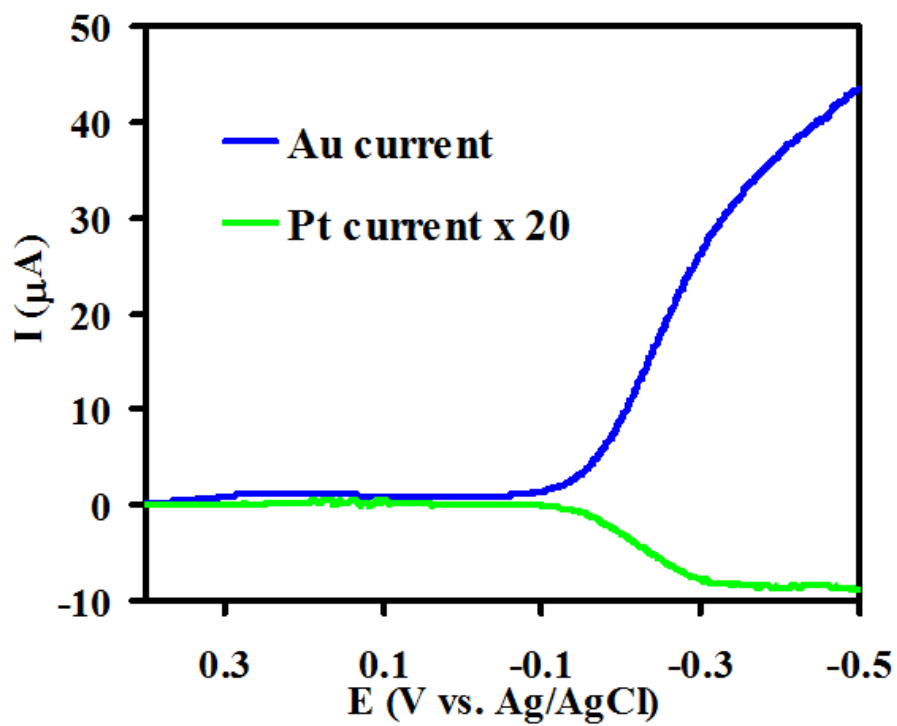

**Figure S8:** RRDE data of FeFc$_4$ physiabsorbed on ATM modified Au electrode, deposited for 10 min showing the disc and Pt ring currents, in air saturated pH 7 buffer at a scan rate of 10 mV/s and rotation speed of 300 rpm, using Ag/AgCl reference and Pt wire counter electrodes.

| Table S1: Comparison of peak splitting for bare Au and ATM modified gold surfaces after different time of immersion using redox couple $K_3[Fe(CN)_6]$ in 0.1 M $KNO_3$ ||
|---|---|
| **Samples** | **Peak splitting (mV)** |
| Bare Au | *150* |
| 10 min immersion | *170* |
| 20 min immersion | *210* |
| 30 min immersion | *260* |
| 40 min immersion | *295* |
| 60 min immersion | *335* |
| 180 min immersion | *404* |
| 400 min immersion | *410* |

**Details of Koutecky-Levich (K-L) analysis:**

The electrocatalytic $O_2$ reduction at different rotation rates (RDE) have been performed for $FeFc_4$ immobilized on ATM modified Au. This technique not only helps in determining the number of electrons involved in ORR but also provides a direct proof of the stability as well as durability of these catalysts on the electrode surfaces (Figure 9A, main text). The $O_2$ reduction current increases with increasing rotation rates following the Koutecky–Levich equation, $I^{-1} = i_K^{-1} + i_L^{-1}$, where $i_K$ is the potential dependent kinetic current and $i_L$ is the Levich current.[1] $i_L$ is expressed as $0.62nFA[O_2](D_{O2})^{2/3}\omega^{1/2}v^{-1/6}$, where $n$ is the number of electrons transferred to the substrate, $A$ is the macroscopic area of the disc (0.125 cm$^2$), [$O_2$] is the concentration of $O_2$ in an air saturated buffer (0.2 mM) at 25 $^o$C, $D_{O2}$ is the diffusion coefficient of $O_2$ (1.8 x 10$^{-5}$ cm$^2$ s$^{-1}$) at 25 $^o$C, $\omega$ is the angular velocity of the disc and $v$ is the kinematic viscosity of the solution (0.009 cm$^2$ s$^{-1}$) at 25 $^o$C.[2] The plot of $I^{-1}$ at multiple rotation rates *vs.* the inverse square root of the angular rotation rate ($\omega^{-1/2}$) is linear. The slopes obtained from the experimental data for $FeFc_4$ closely matches with the theoretical slope predicted for a 4e$^-$ process (Figure 9B, main text).